\pdfoutput=1

\documentclass[11pt,a4paper]{article} 
\usepackage{jcappub} 

\usepackage{amsmath}
\usepackage{amsfonts}
\usepackage{graphicx}
\usepackage{bm}

\usepackage{varioref}
\usepackage{amssymb}
\usepackage{comment}

\title{Exploring a string-like landscape}
\author{Jonathan Frazer}
\author{and Andrew R. Liddle} 
\affiliation{Astronomy Centre, University of Sussex, Brighton BN1 9QH,
United Kingdom}

\emailAdd{J.Frazer@sussex.ac.uk}
\emailAdd{A.Liddle@sussex.ac.uk}

\abstract{
We explore inflationary trajectories within randomly-generated two-dimensional potentials, considered as a toy model of the string landscape. Both the background and perturbation equations are solved numerically, the latter using the two-field formalism of Peterson and Tegmark which fully incorporates the effect of isocurvature perturbations. Sufficient inflation is a rare event, occurring for only roughly one in $10^5$ potentials. For models generating sufficient inflation, we find that the majority of runs satisfy current constraints from WMAP. The scalar spectral index is less than 1 in all runs. The tensor-to-scalar ratio is below the current limit, while typically large enough to be detected by next-generation CMB experiments and perhaps also by Planck. In many cases the inflationary consistency equation is broken by the effect of isocurvature modes.
}

\begin{document}

\maketitle
\flushbottom


\section{Introduction}

Historically researchers have hoped that there is a fundamental physical theory uniquely describing what we observe. More than that, many have hoped that it has a signature allowing us to distinguish it from any other theory we might have imagined along the way. While string theory may or may not be the answer to all our questions, what is exciting is that for the first time it gives us a theory that might be (see ref.~\cite{Burgess} for a nice review of string theory in cosmology). But string theory has surprised those who work with it time and time again. First there was one string theory, then there were five, then it turned out there was a continuum of theories~\cite{HW,Susskind}!

How are we to interpret this? Susskind coined the term {\em landscape} \cite{Susskind}. He described what he called the megaverse (now more commonly referred to as the multiverse), where in the low-energy approximation, different regions may be characterized by the values of a large number of scalar fields. The consequence of this is that we have some very complicated potential $V(\phi_{1},...,\phi_{D})$, with a very large number of minima, each corresponding to a different metastable vacuum energy. This implies that instead of trying to predict the values of observables, we should be trying to predict probability distributions. 

An early study of the possible consequences of this landscape picture for inflation was carried out by Tegmark \cite{TegmarkInf}, who generated a large number of random one-dimensional potentials and explored the inflationary outcomes. However a one-dimensional approach gives a very limited view as compared to the possible dynamics of the landscape, in particular limiting the effect of the choice of initial conditions and restricting the observable outcome to adiabatic perturbations. In this article we take further steps towards a more realistic rendition of the landscape, while remaining quite rudimentary, by carrying out a similar analysis in a two-dimensional field space. This broadens the effect of initial conditions, as there are now a family of possible trajectories passing through each point in field space, and permits isocurvature perturbations which can modify the form of the late-time adiabatic spectrum. Our aim is to characterize the spread of observational predictions for such models. In future work we will extend further to a $D$-dimensional landscape, which adds computational complexity but no new issues of principle, unlike the extension from one to two dimensions.

\section{Approach}

\subsection{Where we intend to explore}

Ideally we would like to explore the landscape potential $V(\phi_{1},...,\phi_{D})$ directly but sadly the explicit form is currently unknown. So to be getting on with we construct an artificial potential. The hope is we can explore how certain characteristics expected to be manifest in the true potential give rise to a probability distribution for observables. Adapting the approach taken in ref.~\cite{TegmarkInf}, we define a potential
\begin{equation}
V(\phi_1,..,\phi_{D})=m_{\rm{v}}^{4}f\left(\frac{\phi_{1}}{m_{\rm{h}_{1}}},..,\frac{\phi_{D}}{m_{\rm{h}_{D}}}\right),
\end{equation}
where $f$ is a well-behaved dimensionless function and $m_{\rm{v}}$ and $m_{\rm{h}_i} $, $(i \in 1..D)$, are the characteristic vertical and horizontal mass scales. Ultimately $m_{\rm{v}}$ is to be adjusted to give the correct amplitude of observed perturbations. Adjusting this mass does not affect anything other than the distribution of vacuum energies, so for now we will not concern ourselves with it. 

In ref.~\cite{TegmarkInf} the case where $D=1$ was extensively investigated. We wish to move to higher $D$ as it allows for a broader range of behaviour, particularly isocurvature perturbations which will be one of the main focuses of this paper. In this paper we only take the step of increasing $D$ from 1 to 2; however we believe this is the biggest step as there are no further qualitative differences to go beyond $D=2$. We reserve analysis of that case to future work.\footnote{An exception may be the distribution of the vacua, which changes in an interesting way both classically and particularly with regard to tunnelling \cite{AE, SSS, Tyetunnel} .} Table~\ref{tab:ndchanges} summarizes some of the changes as the number of scalar fields is increased.

\begin{table}[t]
\centering
  \begin{tabular}{ | p{2.5cm}  | p{3.5cm}  | p{3.5cm}  | p{3.5cm} |}
    \hline
     & Single-field & Two-field & Multi-field \\ \hline
   Motivation & All that inflation asks for & Richer behav\-iour yet still simple & Natural consequence of many fundamental theories\\ \hline
   Surfaces & Maxima, Minima, Slopes & Maxima, Minima, Saddles, Valleys, Ridges &  Maxima, Minima, Saddles, Valleys, Ridges   \\ \hline
    Perturbations & Curvature & Curvature and Isocurvature & Curvature and Isocurvature \\ \hline
    Issues & Requires very flat potentials that may be hard to realize in fundamental theory & Initial Conditions & Initial conditions  \\
    \hline
    \end{tabular}
\caption{Summary of some of the differences between single scalar field models and more scalar fields.}
\label{tab:ndchanges}
\end{table}

To construct our landscape, following an approach similar to ref.~\cite{TegmarkInf} we use a random function of the form
\begin{equation}
f(x,y)  = \sum_{j=1}^m \sum_{k=1}^n \left[a_{j,k} \cos{(j x + k y)}+ b_{j, k} \sin{(j x + k y)}\right]
\end{equation}
where in practice we truncate the series at the values $m=n=5$ and the Fourier coefficients $a_{j,k}$, $b_{j, k}$ are independent Gaussian random variables with zero mean and standard deviation
\begin{equation}
\sigma = e^{-(j^2+ k^2)/2 n},
\end{equation}
With this form, the potentials we simulate are periodic with periodicity scale $2\pi m_{\rm{h}_{i}}$, and we can only expect reasonable results if the evolution spans a distance in the $x$--$y$ plane less than the periodicity of the function. For our choice of the  horizontal mass $m_{\rm{h}_{i}}$, which we are about to discuss, the periodicity is large enough to have no effect.

\begin{figure}[t]
\centering
\includegraphics[width=14cm]{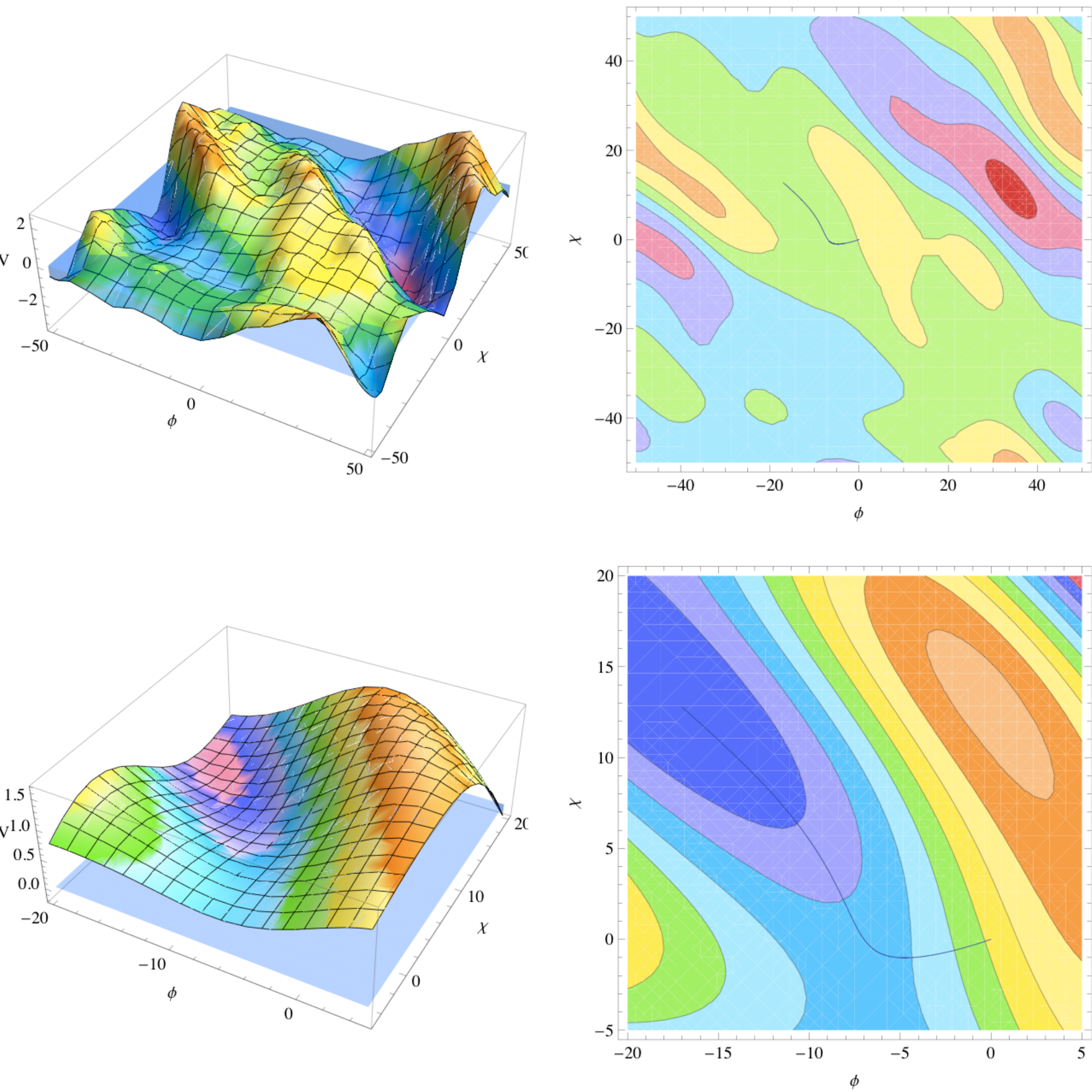}
\caption{An example of our landscape, shown on two scales together with the trajectory taken during inflation. The first scale corresponds to the periodicity of the function, and the second shows the features relevant to the inflationary trajectory.  The ``water level" indicates the height of the minima determining the vacuum energy. We see this is very close to $V=0$; this, together with the fact that this trajectory gives a sufficient number of e-folds of inflation (in this case $N=173$), means that we have an anthropically suitable bubble universe. }
\label{fig:aleverse}
\end{figure}

The precise form of our potential clearly has no theoretical motivation and as such we are free to tinker with it as we please. We are interested in how features of the landscape affect the evolution of the power spectrum and observables, so what we need to consider is on what scale these features occur. If any, it is this quality of our landscape which should be motivated by fundamental theory. Our choice of potential allows us to adjust this in three ways; the truncation number, the standard deviation of the Fourier coefficients, and the choice of $m_{\rm{h}_{i}}$. The truncation number and tuning of the standard deviation are specific to our potential, but the mass scale is a rather more general feature of potentials. Thus, to try and focus on this sense of scale, we fix the deviation and truncation number as stated above and discuss our options in terms of adjusting the horizontal mass with respect to some reference mass, taken to be the reduced Planck mass $M_{\rm{Pl}}$. If we think of features in the landscape as being anything that can cause a change in direction of the inflationary trajectory, then we are essentially asking how many of these features we expect the trajectory to encounter during an evolution giving rise to a sufficiently large number of e-folds of inflation. There are many interesting mass scales that we will not be attempting to explore here, but crudely speaking they fall into four categories:

\begin{description}
  \item[$m_{\rm{h}_{i}}\gg M_{\rm{Pl}}$:] This mass choice is poorly motivated by theory but was nonetheless explored as one of many cases for single scalar field in ref.~\cite{TegmarkInf}. More generally speaking, this case corresponds to a trajectory on a nearly flat, almost featureless potential. The advantage of this is that it is highly predictive since all trajectories look pretty much the same and in the single-field case can readily be in agreement with observation. Also, it is easy to get lots of inflation.
   \item[$m_{\rm{h}_{i}}\ll M_{\rm{Pl}}$:]  This mass choice corresponds to something akin to an egg box.\footnote{Allowing a high truncation number when $m_{\rm{h}_{i}} \gtrsim M_{\rm Pl}$ introduces small-scale power into the potential to similar effect.} The main issue with this choice is that it becomes difficult to achieve sufficient inflation without getting eternal inflation, since the trajectory will almost always roll straight into a minimum. That said, one might imagine that if the number of scalar fields was sufficiently large then the chance of getting sufficient inflation would increase. This is a particularly important effect if tunnelling is taken into consideration \cite{AE, SSS, Tyetunnel},  but to keep things simple we will not be doing so in this paper.
  \item[$m_{\rm{h}_{i}}\ll M_{\rm{Pl}}$ and $m_{\rm{h}_{j}}\gg M_{\rm{Pl}}$:] While our choice of potential would require additional terms to investigate this scenario, we are referring to the sort of case where there is more than one kinematically significant scale. An example of this kind of situation was investigated in refs.~\cite{TyeXuZhang,TyeXu}. The setup they considered could be imagined as a sort of multi-dimensional version of a board with nails stuck in it. The trajectory has a slow-roll drift velocity with what the authors describe as a brownian motion imposed on top of this. Depending on the scales involved, this could lead to interesting features in the power spectra and, for a given suitable background evolution, the extra distance covered due to the random motion will increase the number of e-folds. 
  \item[$m_{\rm{h}_{i}}\sim M_{\rm{Pl}}$:] Finally we have the beginners' ski slope, an example of which is shown in Fig.~\ref{fig:aleverse}. This is what we will be investigating. There are no jumps and the features are gentle so an advanced skier or snowboarder would probably be rather bored in our landscape, but for inflation we feel this is an interesting scale on which to begin our exploration. This mass scale is well motivated by theory and it also gives rise to quite a broad range of behaviour, since in order to obtain a sufficient number of e-folds of inflation, the trajectory generally has to take a non-straight path. The downside is that the broad range of behaviour will make the model less predictive but then again, it is interesting to see how robust the values of certain parameters are under such variations. Also, while the variability in a two scalar field model may be large, one can easily imagine that the deviation might decrease as more scalar fields are introduced. This tendency was seen in ref.~\cite{KL} for Nflation models with random initial conditions on many uncoupled fields.
\end{description}

\subsection{How we explore it}

As we discussed, any model of inflation where the potential has multiple minima predicts a probability distribution for the cosmological parameters. We wish to compute this distribution for the potential described above. To do this we perform essentially the same experiment that was performed in ref.~\cite{TegmarkInf}: 

\begin{enumerate}
\item Generate a random potential $V(\mathbf{\phi})$ and start at $\mathbf{\phi}=(0,0)$.
\item If $V(0,0)<0$ then reject model, otherwise evolve to find the field trajectory.
\item If model gets stuck in eternal inflation, reject.
\item Once the model stops inflating, if the number of e-folds of inflation $N<60$ we reject as insufficient inflation occurred. Otherwise find the local minimum to calculate $\rho_{\rm vac}$, and if  $\rho_{\rm vac}<0$, reject. If $N \ge 60 $ and $\rho_{\rm vac}\ge 0$, calculate observables. 
\item Repeat many times to obtain a statistical sample.
\item (Change some assumptions and do it all again.)
\end{enumerate}

We start the evolution at $V(0,0)$ for practical purposes. Instead of viewing this as starting at the same position in many different potentials, since the statistics of our landscape are invariant under translation, one could equally view this as starting at random positions in one infinite potential. An alternative choice of initial conditions would be to start only at maxima. This relates to the choice of measure, a problem we will discuss shortly but which in reality is an issue well beyond the scope of this paper. Step 2 is there since it would make no sense to do otherwise; generally we are not interested in the statistics of initial conditions that don't give rise to regions of space. Step 3 rejects trajectories that get stuck in minima that lead to eternal inflation. We will not be considering tunnelling in this paper, so once again we give no statistical weighting. 

Step 4 describes what we do when we finally do encounter a potentially viable trajectory. The rejections that take place here are more debatable since they are essentially anthropic rejections. In ref.~\cite{Weinberg} it was shown that anthropic arguments place an incredibly strong bound on the vacuum energy. This was  and still is a very important result and the realization that it sits so naturally with string theory is more than a little exciting. We do not explore this idea in our current work; indeed we ignore it since the extremely narrow permitted range for the present vacuum energy is not computationally accessible. Instead we place only an approximate lower bound, rejecting universes with a negative vacuum energy on the grounds that they will recollapse shortly after inflation ends. Technically very slightly negative vacuum energies are anthropically acceptable but this does not affect what we are interested in here. Rejecting inflation of less than 60 e-folds comes from arguing that this is a requirement for the formation of galaxies on a smooth cosmological background \cite{TegmarkInf,tunland}. Steps 5 and 6 follow trivially.

In constructing our experiment in this way we have inadvertently chosen a measure. We do not wish to go into it in any depth here but it is important to realize that what is known as \emph{The Measure Problem} \cite{TegmarkInf, measure} directly affects this work. If we subscribe to Bayesian statistics then one can describe the current situation as follows:
Define $P(U)$ to be the probability of finding yourself in a particular region of space-time where inflation has ended which we now define as a bubble universe $U$, from now on simply referred to as a universe. Then define $P(O)$ to be the probability of making an observation $O$ in any bubble universe. What one defines as an observation changes the results dramatically. One could simply decide observation meant finding protons for instance. Here, observation corresponds to an observer measuring a particular set of cosmological parameters. The conditional probability of making an observation $O$ in a universe $U$ is $P(O|U)$, called the likelihood. Meanwhile, the probability that we are in a universe $U$ given that we have the observation $O$ is $P(U|O)$, called the posterior probability. Bayes' theorem then gives us
\begin{equation}
P(U|O)=\frac{P(O|U)P(U)}{P(O)}.
\end{equation} 
Our experiment essentially makes an attempt at calculating $P(O)$ but this can be written as
\begin{equation}
P(O)=\int_{M}P(O|U)P(U)dU
\end{equation} 
where $M$ is the multiverse, and $dU$ is the measure. So as you can see we have issues. The probability of making an observation $O$ given $U$ is where anthropic arguments come in.  For example, the probability of making an observation in a universe that can't form galaxies is, we guess, not very high. We also have the problem of counting infinities and this is the measure problem. We have possibly an infinite number of universes, infinite in extent and yet we need to count the probability of making observations in them. We do not wish to discuss this in any detail but we need to remember that our experiment, for the time being, uses a measure where initial conditions are weighted equally (step 1) to start with but then we give zero weighting to universes that are invalid on anthropic grounds (step 4). 

\section{Calculating observables}

In this section we lay down the theoretical framework involved in calculating observables. Much of this section follows ref.~\cite{PTtwofield}, which gives an extremely thorough and comprehensive review of how to make two-field model calculations. We therefore only give a summary of the calculations involved, operating in a simplified setting where we assume the fields are canonically normalized and working to lowest order in the perturbations (since the effect of higher-order terms would be much smaller than the uncertainty of our overall implementation). First we introduce the unperturbed equations. In the context of our work these are the equations used to perform the experiment described in the previous section. Next we move on to the background kinematics. These will prove invaluable in our attempts to understand what aspects of the potential are important in affecting observables and how these characteristics extend to a larger number of scalar fields. We then move on to the perturbed equations which finally lead us to calculating observables. We use reduced Planck units ($M_{\rm Pl}=1$)  throughout and the vector notation $\mathbf{\phi}=(\phi_{1},..,\phi_{D})$. In practice we will often simply have $\mathbf{\phi}=(\phi_{1},\phi_{2})$ but where it will be helpful in later discussion, we write down explicitly how this extends to $D$ scalar fields.

\subsection{Background equations}

We take the background spacetime to be flat, $(3+1)$-dimensional, homogeneous and isotropic and thus described by the Robertson--Walker metric,
\begin{equation}
ds^2=-dt^2+a^2(t)\left[dx^2+dy^2+dz^2\right],
\end{equation}
where $a(t)$ is the scale factor. We are investigating inflation driven by $D$ scalar fields with canonical kinetic terms. The non-gravitational part of the action is
\begin{equation}\label{eq:action}
S=\int \left[-\frac{1}{2}\sum_{i=1}^{D}\partial^{\mu}\phi_{i}\partial_{\mu}\phi_{i}-V(\phi_{1},..,\phi_{D})\right] \sqrt{-g}\,d^4 x,
\end{equation}
and thus we have the standard field equation
\begin{equation}\label{eq:fieldcomovingt}
\ddot{\mathbf{\phi}} + 3H\dot{\mathbf{\phi}} + \nabla^{\dagger} V = 0,
\end{equation}
where $H$ is the Hubble parameter, whose dynamics are found from the $(0,0)$ component of Einstein's equations,
\begin{equation}\label{eq:Hparam}
3H^2=\rho.
\end{equation}
From eq.~\eqref{eq:action} we also obtain 
\begin{equation}
\rho = \frac{1}{2}|\dot{\mathbf{\phi}}|^2+V\quad,\quad P=\frac{1}{2}|\dot{\mathbf{\phi}}|^2-V
\end{equation}
and so eq.~\eqref{eq:Hparam} becomes
\begin{equation}\label{eq:Hcomovingt}
3H^2=\frac{1}{2}|\dot{\mathbf{\phi}}|^2+V.
\end{equation}

Working in terms of comoving time $t$ is not particularly convenient for our current work. Instead it is much more helpful to make the transformation
\begin{equation}\label{eq:ttrans}
dN\equiv Hdt,
\end{equation}
where $N$ is the logarithmic growth of the scale factor and represents the number of e-foldings of inflation. Using $N$ is useful since it is directly related to observables; it also simplifies all the equations and renders them dimensionless, thus making it simpler to compare the relative size of terms. From now on we will use the notation
\begin{equation}
'\equiv\frac{d}{dN}
\end{equation}
to represent differentiation with respect to $N$. 

To further simplify the equations we introduce the slow-roll parameter
\begin{equation}\label{eq:epsilon}
\epsilon\equiv-\frac{\dot{H}}{H^2}.
\end{equation}
Inflation ends when $\ddot{a}\leq0\iff\epsilon\geq1$. If we combine the field equation \eqref{eq:fieldcomovingt}  with the comoving time derivative of eq.~\eqref{eq:Hcomovingt} we see that the slow-roll parameter can also be interpreted in terms of the field speed;
\begin{equation}
v\equiv|\mathbf{\phi}'|=\sqrt{2\epsilon}.
\end{equation}
Making the transformation given in eq.~\eqref{eq:ttrans} and substituting eq.~\eqref{eq:epsilon}, the field and Friedman equations simplify to
\begin{equation}\label{eq:field}
\mathbf{\phi}'+\nabla^{\dagger}\ln V=-\frac{\mathbf{\phi}''}{3-\epsilon}
\end{equation}
and
\begin{equation}
H^2=\frac{V}{3-\epsilon}
\end{equation}
respectively.

Equation~\eqref{eq:field} together with eq.~\eqref{eq:epsilon} give the evolution of the fields. To solve them we need to provide initial conditions. We choose these to be $\mathbf{\phi}=(0,0)$ and $\mathbf{\phi}'=(-V,_{\phi_{1}}/V,-V,_{\phi_{2}}/V)$. 

\subsection{Relating kinematics to the potential via the slow-roll slow-turn approximations}

This subsection is predominantly a summary of ref.~\cite{ PTtwofield}, although our case is simpler in that we are only concerned with canonical kinetic terms. We also make the extension to a larger number of scalar fields a little more explicit.

We wish eventually to investigate the evolution of perturbations. As summarized in Table~\ref{tab:ndchanges}, when there is only one scalar field, the only type of perturbation that exists is density perturbations, generated by perturbing the field along the trajectory. When there are more scalar fields however, for every extra scalar field there is another direction perpendicular to the trajectory in which perturbations can also arise. The perturbations perpendicular to the trajectory correspond to isocurvature perturbations but these in turn can fuel the evolution of the density perturbations. Decomposing the perturbations in this way leads us to introduce a new basis. If we think of the old basis as $B=\{e_{1},..,e_{D}\}$ where $e_{1}$ points in the direction $\phi_{1}$, $e_{2}$ points in the direction $\phi_{2}$ and so on, then we can think of the new basis $K=\{e_{\parallel},e_{\perp_{1}},..,e_{\perp_{D-1}}\}$ as a rotation of the old basis such that the first basis vector now points along the trajectory and we label it $e_{\parallel}$. The other basis vectors will then point perpendicular to the trajectory and we label them $e_{\perp_{i}}$. We refer to this as the \emph{kinematic basis}\footnote{For a fully kinematic basis, one could also define one of the perpendicular fields such that the direction of the turning of the field is entirely along it, but we choose not to do that here since our basis will make it clearer how to extend our statistics to $D$ fields.}  and we denote the components of a general vector $\mathbf{A}$ and matrix $M$ as, for instance 
\begin{equation}
A_{\parallel}=e_{\parallel}\cdot\mathbf{A}\quad,\quad M_{\parallel\perp_{i}}=e^{\dagger}_{\parallel}M e_{\perp_{i}}.
\end{equation}

Using the kinematic basis we can now introduce the slow-roll slow-turn (SRST) approximations. There is more than one way to interpret the well-known single-field slow-roll approximations and as discussed in ref.~\cite{PTtwofield}, this affects how the slow-roll approximations generalize to multiple scalar fields. With regard to how slow-roll extends, probably the minimal requirement is to say:
\begin{itemize}
\item Expansion is nearly exponential.
\item Deviation from this expansion changes slowly.
\end{itemize}
If we take this as our definition then our slow-roll conditions become
\begin{equation}
\epsilon=\frac{1}{2}v^2\ll 1\quad,\quad\left|\frac{\phi_{\parallel}''}{v}\right|\ll1.
\end{equation}
Although these are sufficient to guarantee the above requirements, it is helpful to have a slow-turn approximation which we define as
\begin{equation}
\left|\frac{\phi_{\perp_{i}}''}{v}\right|\ll1. 
\end{equation}
There are $D-1$ of these slow-turn conditions and violation of any one of these conditions would render the trajectory no longer slowly turning, so one might prefer to instead write
\begin{equation}
\frac{1}{D-1}\left|\frac{\mathbf{\phi_{\perp}}''}{v}\right|\ll1,
\end{equation}
but for our purpose we find the former more useful. If the field is both slowly rolling and slowly turning we say it satisfies SRST.

Returning to eq.~\eqref{eq:field} we see that the left-hand side represents deviations from the  SRST limit and so under SRST conditions we can write
\begin{equation}\label{eq:fieldapprox}
\mathbf{\phi'}\simeq-\nabla^{\dagger}\ln V
\end{equation}
and the Friedman equation becomes
\begin{equation}
3H^2\simeq V.
\end{equation}

When trying to understand how various features of the potential affect observables, sometimes it is nice to interpret things in terms of the kinematics of the scalar field and at other times perhaps it is better to look directly at the underlying geometry. For this reason, amongst others, it is helpful to be able to approximately jump from one approach to the other. Direct from eq.~\eqref{eq:fieldapprox} we have
\begin{equation}
\epsilon\simeq\frac{1}{2}|\nabla^{\dagger} \ln V|^2
\end{equation}
and differentiating eq.~\eqref{eq:fieldapprox} one obtains
\begin{equation}
\mathbf{\phi''}\simeq-M\nabla^{\dagger}\ln V,
\end{equation}
where $M$ is the Hessian of $\ln V$, otherwise known as the mass matrix,
\begin{equation}\label{eq:M}
M\equiv\nabla^{\dagger}\nabla\ln V.
\end{equation}
We therefore arrive at approximations relating the kinematic quantities directly to the potential:
\begin{equation}
\left(\frac{\phi_{\parallel}''}{v}\right)\simeq-M_{\parallel\parallel},
\end{equation}
termed the speed-up rate, and
\begin{equation}\label{eq:turnapprox}
\left(\frac{\phi_{\perp_{i}}''}{v}\right)\simeq\left|M_{\parallel\perp_{i}}\right|,
\end{equation}
known as the $i$-th component turn rate.

We don't look at the second-order equations here but we refer the reader to ref.~\cite{PTtwofield} should they need them.

\subsection{The perturbed equations}

Continuing to adopt the same approach as that chosen in ref.~\cite{PTtwofield} it can be shown that by using a multifield version of the Mukhanov--Sasaki variable \cite{Mukhanov}, the field perturbations decouple from perturbations in the metric and thus we are able to focus solely on the curvature and isocurvature perturbations. The evolution of the field perturbations can be found by perturbing the equation of motion for the background. The standard result in Fourier space is
\begin{equation}
\delta\ddot{\mathbf{\phi}} + 3H\delta\dot{\mathbf{\phi}}+\left(\frac{k}{a}\right)^2\delta\mathbf{\phi} =
 -\left[\nabla^{\dagger}\nabla V
  - \left(3-\frac{\dot{H}}{H^2}\right)\dot{\mathbf{\phi}}\dot{\mathbf{\phi}}^{\dagger} -\frac{1}{H}\ddot{\mathbf{\phi}}\dot{\mathbf{\phi}}^{\dagger}-\frac{1}{H}\dot{\mathbf{\phi}}\ddot{\mathbf{\phi}}^{\dagger}\right]\delta \mathbf{\phi},
\end{equation}
where $k$ is the comoving wavenumber. Substituting Eqs.~\eqref{eq:ttrans}, \eqref{eq:epsilon} and \eqref{eq:field} one eventually reaches
\begin{equation}\label{eq:fieldpert}
\frac{1}{3-\epsilon}\delta\mathbf{\phi''}+ \delta\mathbf{\phi'}+\left(\frac{k^2}{a^2 V}\right)\delta\mathbf{\phi}=-\left[M+\frac{\mathbf{\phi''}\mathbf{\phi''}^{\dagger}}{(3-\epsilon)^2}\right]\delta\mathbf{\phi}
\end{equation}
and we see that mode evolution is primarily governed by the mass matrix with small corrections. 

Rotating to the kinematic basis, we can separate eq.~\eqref{eq:fieldpert} to find evolution equations for the adiabatic ($\delta\phi_{\parallel}$) and entropy modes ($\delta\phi_{\perp}$). Starting with the adiabatic mode, projecting in the $e_{\parallel}$ direction, then solving the resulting equation in the super-horizon limit $\left(k/aH\right)^2\ll1$, one finds that the growing super-horizon adiabatic modes are described by 
\begin{equation}\label{eq:adiabaticpert}
\delta\phi_{\parallel}'= \left(\frac{\phi_{\parallel}''}{v}\right)\delta\phi_{\parallel}+2\sum_{i}^{D-1}\left(\frac{\phi_{\perp_{i}}''}{v}\right)\delta\phi_{\perp_{i}},
\end{equation}
or equivalently 
\begin{equation}\label{eq:speedupevol}
\left(\frac{\delta\phi_{\parallel}}{v}\right)'= +2\sum_{i}^{D-1}\left(\frac{\phi_{\perp_{i}}''}{v^2}\right)\delta\phi_{\perp_{i}},
\end{equation}
or when SRST holds
\begin{equation}
\delta\phi_{\parallel}'\simeq M_{\parallel\parallel}\delta\phi_{\parallel}-2\sum_{i}^{D-1}M_{\parallel\perp_{i}}\delta\phi_{\perp_{i}}.
\end{equation}
So we see that the evolution of adiabatic modes can be inferred directly from the background kinematics. The first term of eq.~\eqref{eq:adiabaticpert} tells us that the faster the speed-up rate, the faster the evolution of the mode. The second term tells us about the sourcing of the adiabatic modes from the entropy modes. We see that for a given size of entropy mode, the faster the turn rate, the more the adiabatic mode will be sourced by that entropy mode. For our experiment we will see there is a distribution associated with the likelihood of encountering a given turn rate. This means that as we extend to a larger number of scalar fields, it becomes increasingly likely that the fuelling from entropy modes will be a significant effect. 

Following the same procedure for the entropy modes we find that
\begin{equation}\label{eq:entropypert}
\frac{\delta\phi_{\perp_{i}}''}{3-\epsilon}+\delta\phi_{\perp_{i}}'=-\mu_{\perp_{i}}\frac{\delta\phi_{\perp_{i}}}{v},
\end{equation}
where $\mu_{i}$ is the effective entropy mass
\begin{equation}
\mu_{\perp_{i}}\equiv M_{\perp_{i}\perp_{i}}+\frac{9-\epsilon}{(3-\epsilon)^2}\left(\frac{\phi_{\perp_{i}}''}{v}\right)^2
\end{equation}
and so is well approximated by $\mu_{\perp_{i}}\simeq M_{\perp_{i}\perp_{i}}$. The great news here is that eq.~\eqref{eq:entropypert} tells us that each of the entropy modes evolves independently of the others. This means we can find their amplitude and determine the evolution of adiabatic modes without solving a massive set of fully coupled equations. We also see that the evolution of a given entropy mode is dominated by the curvature of the $\log$ of the potential along that entropic direction. This makes sense intuitively as when the curvature is positive, a given trajectory is stable in the sense that a perturbed trajectory will be redirected back onto the background trajectory. Conversely, when the curvature is negative, the trajectory is dispersive and perturbations will evolve, fuelling the growth of that entropy mode.

From here we are in a position to discuss curvature and isocurvature perturbations. The curvature perturbation represents the perturbation in the curvature of constant-time hypersurfaces. In the comoving gauge we have the gauge-invariant quantity, $R$, which during inflation can be shown to be
\begin{equation}
R=\frac{\delta\phi_{\parallel}}{v}.
\end{equation}
Isocurvature perturbations represent relative fluctuations in the different fields that leave the total curvature unchanged and hence are related to entropy perturbations, here defined as\footnote{This definition, following refs.~\cite{WBMR,PTtwofield}, is chosen so that the isocurvature mode typically has the same amplitude as the adiabatic one at horizon crossing, and differs from the alternative convention where the isocurvature perturbations are normalized like massless field perturbations.}
\begin{equation}
S_{i}\equiv\frac{\delta\phi_{\perp_{i}}}{v}.
\end{equation}
Using eq.~\eqref{eq:speedupevol} then gives us
\begin{equation}\label{eq:R'}
R'= +2\sum_{i}^{D-1}\left(\frac{\phi_{\perp_{i}}''}{v}\right)S_{i}
\end{equation}
and so we see that the super-horizon evolution of the density perturbation is independent of the speed-up rate; instead it depends on the sum of the isocurvature perturbations, each multiplied by their corresponding turn rate. We also see that when there is no turn rate (in any of the directions), the single-field result that the density perturbation is conserved on super-horizon scales is recovered. 


For the isocurvature perturbations we use the approach of ref.~\cite{WBMR} and parameterize the isocurvature modes as 
\begin{equation}\label{eq:S'}
S_{i}'=\beta_{i} S_{i}.
\end{equation}
We then find that $\beta$ can be well approximated in the SRST limit as
\begin{equation}
\beta_{i}\simeq M_{\parallel\parallel}-M_{\perp_{i}\perp_{i}}
\end{equation}
So interpreting this from the geometrical point of view we see that the evolution of a given set of isocurvature modes depends on the integral of the difference between the curvature along the adiabatic and entropic directions in a very intuitive manner. As summarized in Table~\ref{tab:isoevol}, negative curvature in the adiabatic direction and positive curvature in the entropic direction corresponds to maximum damping of isocurvature modes, while positive curvature in the adiabatic direction and negative curvature in the entropic direction corresponds to maximum fuelling.

\begin{table}[t]
\centering
\begin{tabular}{|c|c|c|}
\hline
& $M_{\parallel\parallel}>0$&  $M_{\parallel\parallel}<0$\\
\hline 
 $M_{\perp\perp}>0$ & Playoff & Damping\\
 \hline
 $M_{\perp\perp}<0$& Fueling & Playoff \\
\hline
\end{tabular}
\caption{Origins of evolution of isocurvature modes.}
\label{tab:isoevol}
\end{table}

Plugging eq.~\eqref{eq:S'} into eq.~\eqref{eq:R'} we have the result
\begin{equation}\label{eq:R}
R\simeq R_{*}+\sum_{i}^{D-1}\int_{N_{*}}^{N}2\frac{\phi_{\perp_{i}}''}{v}e^{\int_{N_{*}}^{\tilde{N}}(M_{\parallel\parallel}-M_{\perp_{i}\perp_{i}})d\tilde{\tilde{N}}}d\tilde{N}.
\end{equation}
and the total isocurvature perturbation $S$ is given by
\begin{equation}\label{eq:S}
S=\sum_{i}^{D-1}S_{i}\simeq \sum_{i}^{D-1}S_{*i}e^{\int_{N_{*}}^{N}(M_{\parallel\parallel}-M_{\perp_{i}\perp_{i}})d\tilde{N}},
\end{equation}
where $*$ denotes that a given quantity is to be evaluated at horizon crossing. Continuing with the approach of ref.~\cite{WBMR} it will be helpful to rewrite eqs.~\eqref{eq:R} and \eqref{eq:S} as 
\begin{equation}\label{eq:transfermatrix}
\begin{pmatrix}
 R \\
 S_{1}\\
 \vdots\\
 S_{D-1} 
\end{pmatrix}=\begin{pmatrix}
 1 & T_{RS_{1}} & \cdots & T_{RS_{D-1}}  \\
 0 &  T_{SS_{1}}  \\
  \vdots &  & \ddots \\
 0 & & & T_{SS_{D-1}} 
\end{pmatrix}
\begin{pmatrix}
 R_{*} \\
 S_{1*}\\
 \vdots\\
 S_{D-1*} 
\end{pmatrix}
\end{equation}
where the transfer functions $T_{RS_{i}}$ and $T_{SS_{i}}$ are given by 
\begin{equation}\label{eq:TRS}
T_{RS_{i}}(N_{*},N)\equiv \int_{N_{*}}^{N}2\frac{\phi_{\perp_{i}}''}{v}T_{SS_{i}}(N_{*},\tilde{N})d\tilde{N}.
\end{equation}
and
\begin{equation}\label{eq:TSS}
T_{SS_{i}}(N_{*},N)\equiv e^{\int_{N_{*}}^{N}(M_{\parallel\parallel}-M_{\perp_{i}\perp_{i}})d\tilde{N}}
\end{equation}
respectively. To reduce computational effort we substitute the approximation of eq.~\eqref{eq:turnapprox} when calculating $T_{RS}$ \eqref{eq:TRS}. In the cases we tested we found this approximation to be more than satisfactory.

\subsection{Power spectra and cosmological parameters}

We are now finally ready to discuss the evolution of the power spectra. For a general quantity $\mathcal{X}$, the power spectra and cross spectra are defined as 
\begin{equation}
P_{\mathcal{X}}\delta^{3}(\mathbf{k-\tilde{k}})\equiv \frac{k^3}{2\pi^2}\langle\mathcal{X}(\mathbf{k}),\mathcal{X}^{\dagger}(\mathbf{\tilde{k}})\rangle
\end{equation}
and
\begin{equation}
C_{\mathcal{X}\mathcal{Y}}\delta^{3}(\mathbf{k-\tilde{k}})\equiv \frac{k^3}{2\pi^2}\langle\mathcal{X}(\mathbf{k}),\mathcal{Y}^{\dagger}(\mathbf{\tilde{k}})\rangle
\end{equation}
respectively. We do not go into details here but quantizing and solving the perturbed equations (see e.g.\ ref.~\cite{PTtwofield}) leads us to
\begin{equation}\label{eq:densitypower}
P_{R_{*}}\simeq \left(\frac{H_{*}}{2\pi}\right)^2\frac{1}{2\epsilon_{*}}\left[1+2(C-1)\epsilon-2CM_{\parallel\parallel}\right]_{*},
\end{equation}
\begin{equation}
P_{S_{i*}}\simeq\left(\frac{H_{*}}{2\pi}\right)^2\frac{1}{2\epsilon_{*}}\left[1+2(C-1)\epsilon-2CM_{\perp\perp}\right]_{*}
\end{equation}
and 
\begin{equation}
C_{RS_{i*}}\simeq\left(\frac{H_{*}}{2\pi}\right)^2\frac{1}{2\epsilon_{*}}\left[-2 C M_{\parallel\perp_{i}}\right]_{*}
\end{equation}
where $C=2-\ln 2-\gamma\approx0.7296$ and $\gamma$ is the Euler--Mascheroni constant. For future use we also introduce the tensor power spectrum which takes the usual form
\begin{equation}\label{eq:tensorpower}
P_{T_{*}}\simeq 8\left(\frac{H_{*}}{2\pi}\right)^2\left[1+2(C-1)\epsilon\right]_{*},
\end{equation}
which is conserved for super-horizon modes. We used the above second-order expressions in our computation but as one would expect, there would have been almost no difference had we used first-order approximations.

Applying the transfer matrix eq.~\eqref{eq:transfermatrix} we finally arrive at one of our most important destinations, namely the power spectra at the end of inflation
\begin{equation}\label{eq:PR}
P_{R}=P_{R_{*}}+\sum_{i}^{D-1}\left(2T_{RS_{i}}C_{RS_{i*}}+T_{RS_{i}}^2 P_{S_{i*}}\right),
\end{equation}
\begin{equation}
P_{S}=\sum^{D-1}_{i} P_{S_{i}}=\sum^{D-1}_{i} T_{SS_{i}}^2 P_{S_{i*}}
\end{equation}
and
\begin{equation}
C_{RS_{i}}=T_{SS_{i}}C_{RS_{i*}}+T_{RS_{i}}T_{SS_{i}}P_{S_{i*}}.
\end{equation}
We see most of our previous comments on super-horizon evolution apply in much the same way to the power spectra. It is now also easy to see one of the reasons why we chose to break our exploration down into two stages. The interesting super-horizon behaviour predominantly comes from the transfer functions. We have seen that the super-horizon modes associated with entropy perturbations evolve independently of one another, each separately affecting the evolution of the adiabatic power spectrum. This means that by making a statistical analysis of one set of entropy perturbations, we can easily see how the statistics will generalize to a larger number of scalar fields.

Now that we finally have the power spectra the hard work is over. From here it is straightforward to find approximations for what we consider the key observables, namely the tensor-to-scalar ratio, spectral indices, and the running. We will evaluate these 55 e-foldings before the end of inflation, taken to be the time that the observed scales crossed the horizon during inflation. We call this the pivot scale $N_{\rm pivot}$. All observable quantities are henceforth assumed evaluated at this scale.

The tensor-to-scalar ratio $r$ is defined as the ratio of the tensor power spectrum eq.~\eqref{eq:tensorpower} to the scalar (curvature) power spectrum eq.~\eqref{eq:PR}
\begin{equation}\label{eq:r}
r \equiv \frac{P_{T}}{P_{R}}.
\end{equation}
Note that since the tensor power spectrum is conserved on super-horizon scales, the single-field result provides an upper bound for the multi-field case. We define the spectral index of a power spectrum $P_{\mathcal{X}}$ as
\begin{equation}
n_{\mathcal{X}}\equiv \frac{d \ln P_{\mathcal{X}}}{d \ln k}.
\end{equation}
Note however that the common definition of the scalar spectral index is related to this definition by
\begin{equation}
n_{\rm{s}}=1+n_{R}.
\end{equation}
We find calculating $T_{RS}$ to be computationally demanding so in practice when calculating the spectral index we do the following:
\begin{eqnarray}
\frac{d \ln P_{\mathcal{X}}}{d \ln k} &\simeq & \frac{\ln P_{\mathcal{X}}(N_{{\rm pivot}}+\frac{\Delta N}{2})-\ln P_{\mathcal{X}}(N_{{\rm pivot}}-\frac{\Delta N}{2})}{\Delta N} ; \nonumber\\ 
&\simeq &\ln \frac{P_{\mathcal{X}}(N_{{\rm pivot}})+P_{\mathcal{X}}(N_{{\rm pivot}}+1)}{P_{\mathcal{X}}(N_{{\rm pivot}}-1)+P_{\mathcal{X}}(N_{{\rm pivot}})}.
\end{eqnarray}

The running of the spectral index $\alpha_{\mathcal{X}}$ is a straightforward extension of the spectral index, defined as
\begin{equation}
\alpha_{\mathcal{X}}\equiv \frac{d n_{\mathcal{X}}}{d \ln k},
\end{equation}
and hence we approximate it using the same technique,
\begin{equation}
\frac{d n_{\mathcal{X}}}{d \ln k}\simeq n_{\mathcal{X}}\left(N_{{\rm pivot}}+1/2\right)-n_{\mathcal{X}}\left(N_{{\rm pivot}}-1/2\right).
\end{equation}
The minimum number of computations of $T_{RS}$ required to obtain the running in this way is three, which is why we take averages when calculating the spectral index.

We do not consider non-gaussianity in this paper. A methodology for computing it within the same formalism has now been given in ref.~\cite{PT2}; however, as for instance shown in that paper, it would typically be expected to be small in these types of models. 

\section{Findings}

\begin{table}[t]
\centering
\begin{tabular}{|l|c|c|c|c|c|}
\hline
Quantity & Result &  Observed & Agreement \\
\hline 
$n_{R}$ & $-0.06\pm0.02$ & $-0.027\pm0.014$& Y \\
$\alpha_{R}$ & $-0.0003\pm0.0009$ & $-0.022\pm0.020$& Y \\
$n_{\rm{iso}}$ & $0.001\pm0.13$ & N/A & N/A \\
$\alpha_{\rm{iso}}$ & $ -0.02\pm0.22 $ & N/A & N/A \\
$r$ & $0.05\pm 0.03$ & $ < 0.24 \;(95\% \; \mathrm{c.l.})$ & Y\\
$T_{SS}$ & $0.06\pm0.43$ & N/A & N/A \\
$T_{RS}$ & $0.8\pm0.9$ & N/A & N/A \\
\hline
\end{tabular}
\caption{Some cosmological parameter constraints.}
\label{tab:results}
\end{table}

\begin{figure}[t]
\centering
\includegraphics[width=15cm]{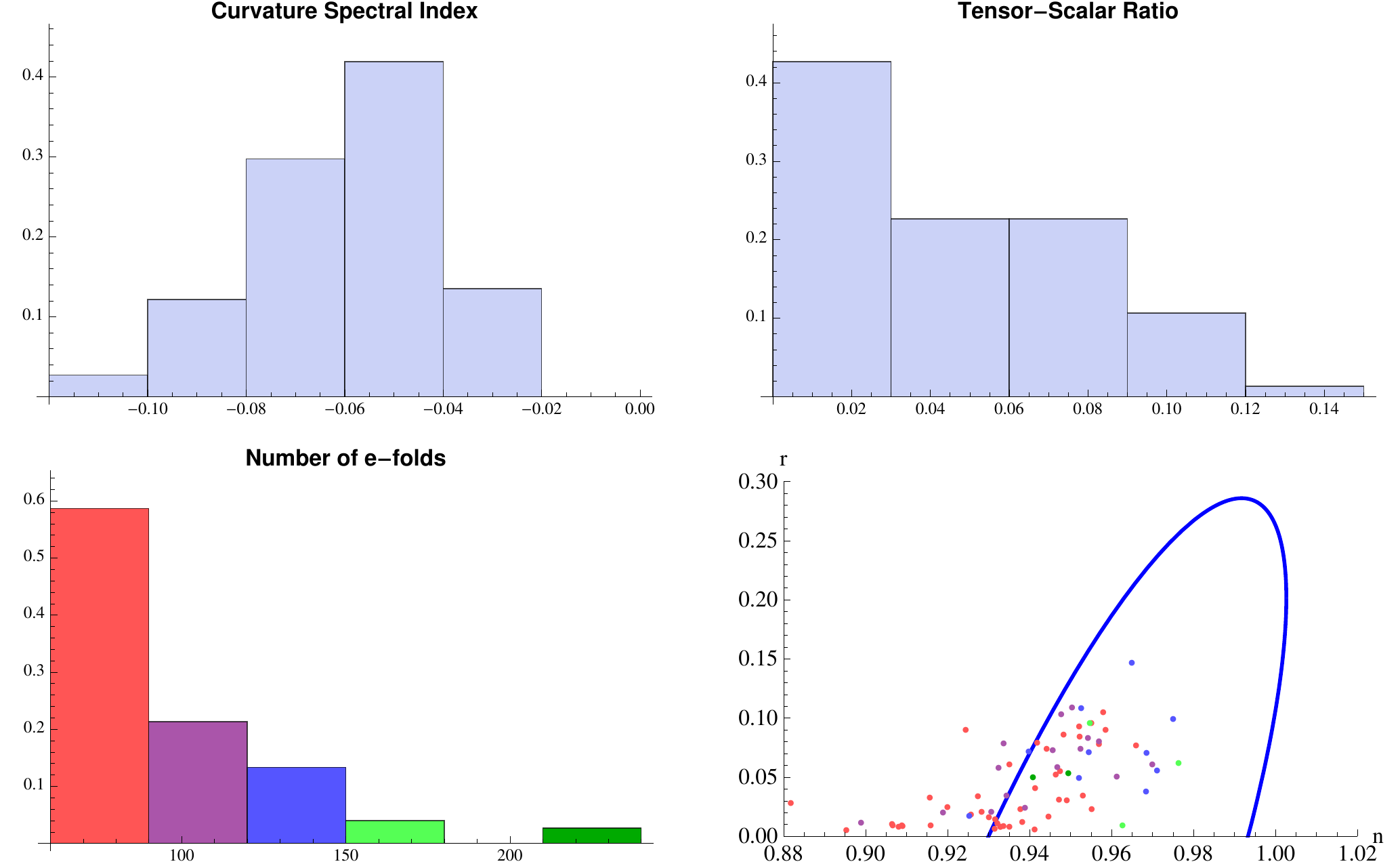}
\caption{Histograms of the curvature spectral index, tensor-to-scalar ratio, and e-folds, plus a plot of the tensor-to-scalar ratio against the spectral index where the points have been colour-coded according to the number of e-folds. The blue line on the $n_{\rm{s}}$--$r$ plot is the $95\%$ confidence contour from WMAP+BAO+$H_{0}$.}
\label{fig:indexplots}
\end{figure}

\subsection{Two scalar fields}

We performed $5\times 10^{6}$ runs and obtained $75$ successful outcomes in terms of sufficient inflation achieved without subsequent collapse. Table~\ref{tab:results} summarizes some of the mean values accumulated and shows that our model is in good agreement with observation in all parameters so far tested. Figure~\ref{fig:indexplots} shows histograms of the curvature spectral index and tensor-to-scalar ratio, a histogram of the number of e-folds and a plot of $r$ against $n_{\rm{s}}$, colour-coded with the number of e-folds. The last plot also shows the present observational limits from a data compilation including WMAP7 results \cite{WMAP7}.

Planck hopes to measure the tensor-to-scalar ratio with an accuracy of a few hundredths, hence has discovery potential if it is of order 0.1 or so. For this reason it is interesting to note that of the universes not already rejected by the $n_{\rm{s}}$--$r$ plot, there seems to be a preference for universes with larger $r$. We also note that all large e-fold universes (green points) lie within the $95\%$ confidence limit and only a couple of blue points are rejected, but more data are needed to conclude whether or not large e-fold universes are favoured within our model. We find this potential trend quite intriguing as it is easy to conceive of measures that give strong weighting to universes with a large number of e-folds. Yet it is generally thought that measures exhibiting this kind of favouritism are rejected observationally due to a tendency to suffer from what is known as the Q-catastrophe (see ref. \cite{gv} for a list of generic problems in constructing measures). Our results hint at a tension with this view, since we find universes with a large amount of inflation can still agree with observation. Our work does not address this problem but it would be interesting to see if this trend is maintained for the very rare cases of universes with a huge number of e-foldings, say in the thousands.

We need to clarify what \emph{agreement} actually means in this context. Both the observed data and the results of our experiment are given in terms of a distribution of values for the observables. As it happens, the statistical spread of each is comparable at present, with a substantial area of overlap; one can therefore conclude that there is a good chance of a single realization from our model giving predictions in accord with the observations. If a model data point lies outside the range observed, then it simply indicates that we do not live in that universe and such data would not necessarily act against the model. Future observations hope to home in on a single value for each observable to ever increasing accuracy, which will clearly soon have higher precision than our model predictions. Provided the observationally-favoured region remains within the envelope predicted by the model, however, such higher-precision measurements will not in themselves be able to argue against the model, at least without further model refinements for instance around the choice of measure and anthropic arguments. At this point we then must invoke the Copernican principle over this remaining anthropically reduced landscape and say we expect to observe values according to the statistics of the model distribution. It is in this sense that we say our results are in agreement with observation. 

The histogram of e-folds, Fig.~\ref{fig:indexplots}, shows that the number of universes drops off rapidly with the number of e-folds. We rejected universes with less than 60 e-folds of inflation but in the process of doing so we found that sufficient inflation was a very rare process, testing millions of universes to find tens of candidates. This is not a new result for ``stringy'' models. For instance ref.~\cite{delicatebrane} investigated one of the most rigorously derived inflationary models from string theory, namely brane--anti-brane inflation, and found much the same thing, while ref.~\cite{tunland} found the same result for tunnelling landscape models. For our model though this is of no concern at all. We are not worried about what proportion of field space allows for anthropically suitable conditions, only that there exists \emph{some} proportion.

\begin{figure}[t]
\centering
\includegraphics[width=14cm]{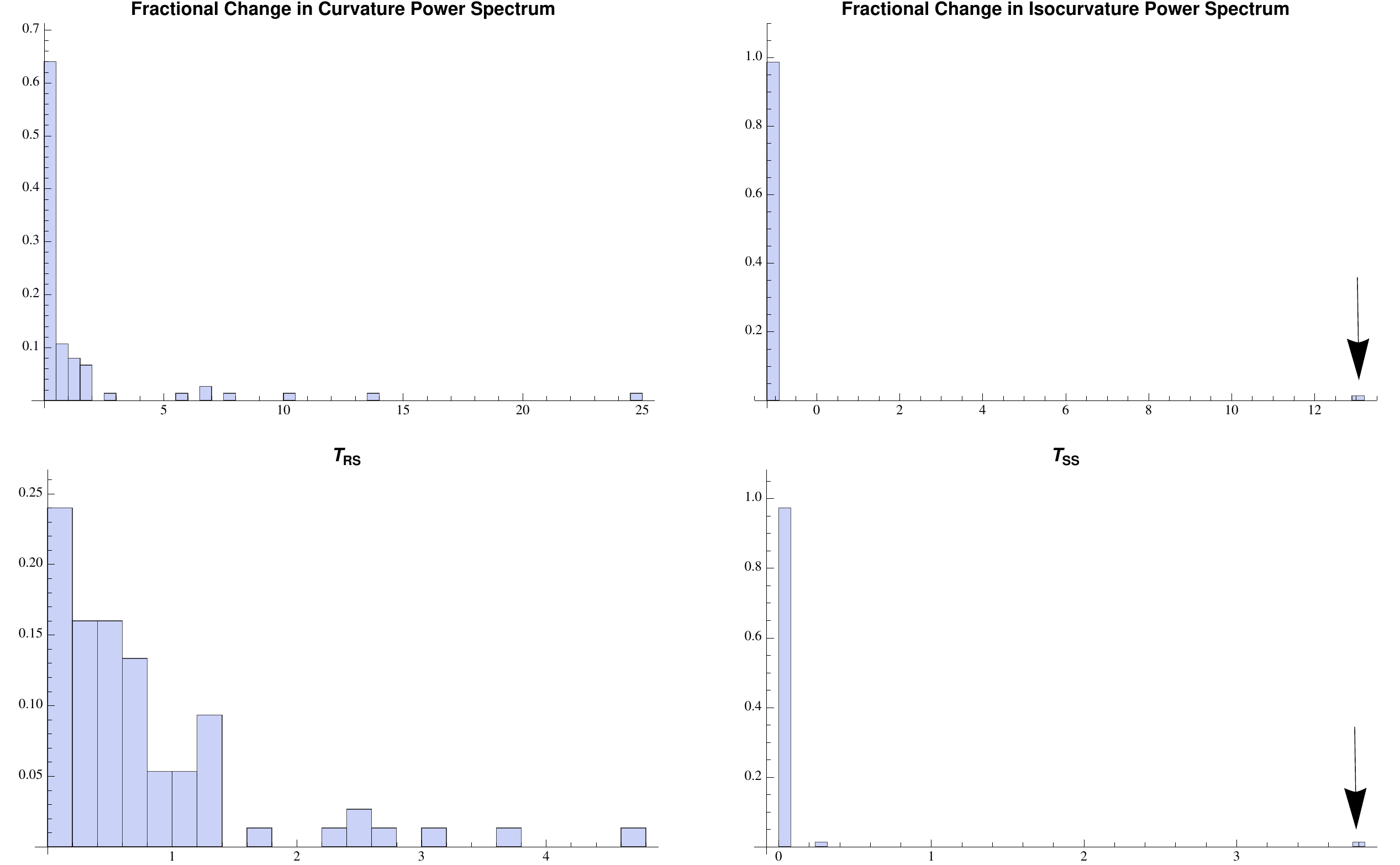}
\caption{Histograms of the fractional change in the curvature and isocurvature power spectra accompanied by histograms for the transfer functions. Arrows indicate Universe 1832942.}
\label{fig:powerstransfers}
\end{figure}

\begin{figure}
\centering
\includegraphics[width=8cm]{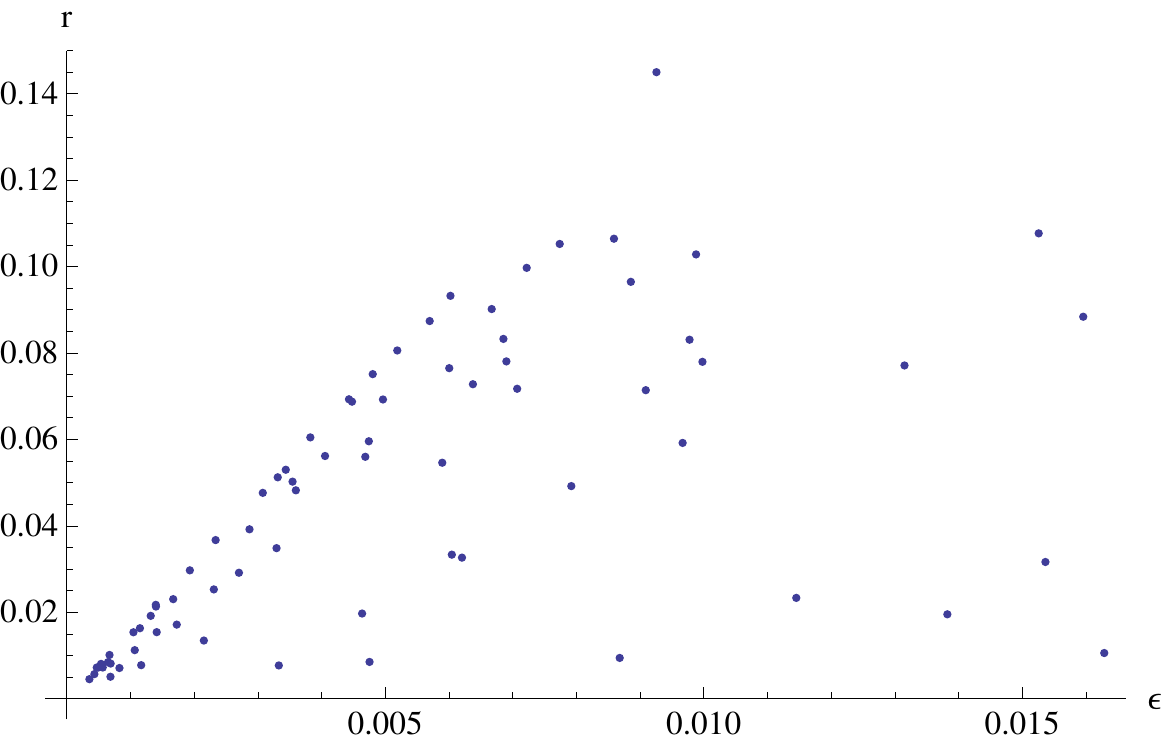}
\caption{Plot of $r$ against $\epsilon$, demonstrating the consistency relation $r=-8n_{\rm T} = 16\epsilon$ becoming an inequality.}
\label{fig:repsilon}
\end{figure}

\begin{figure}[t]
\centering
\includegraphics[width=14cm]{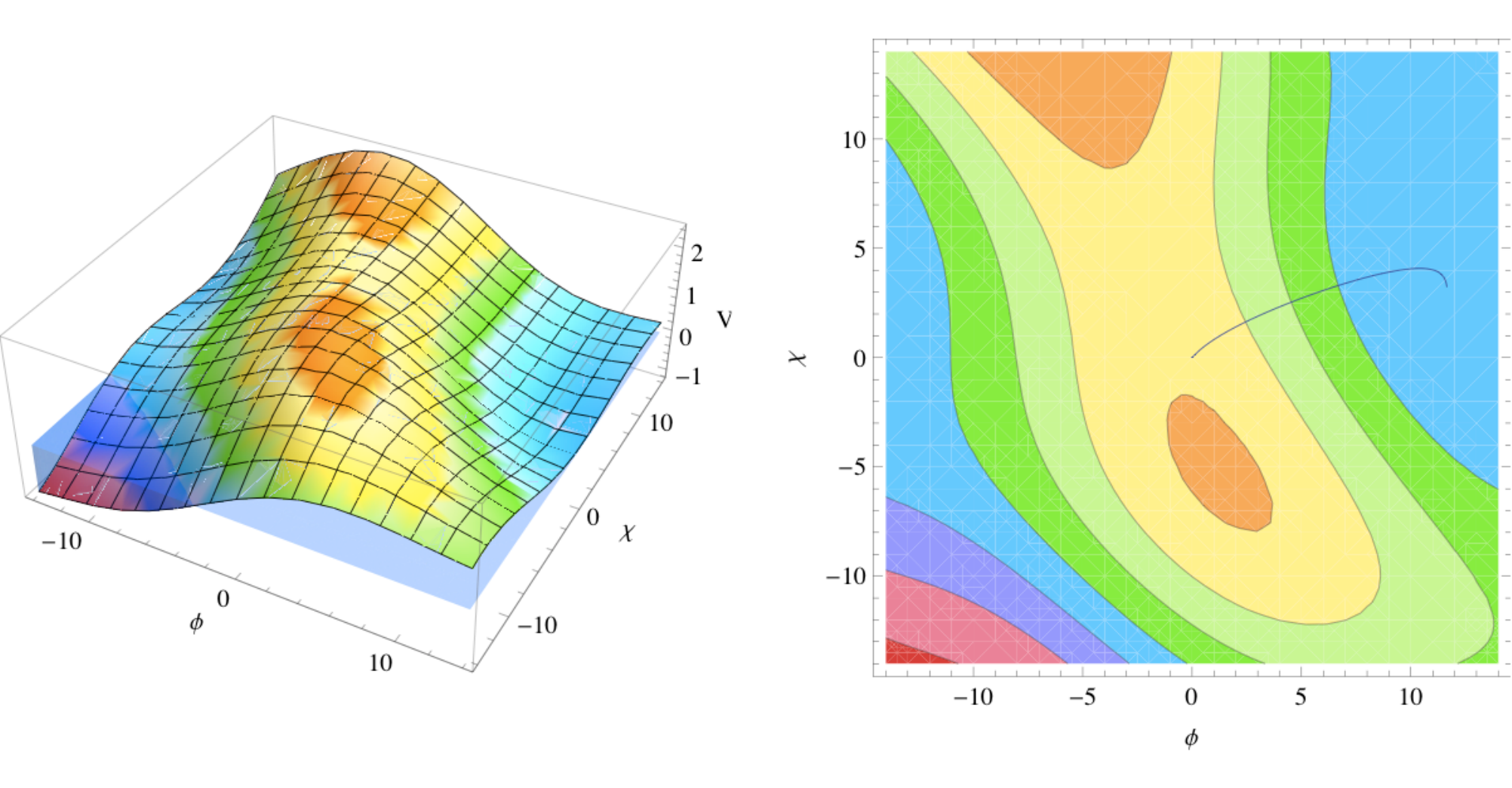}
\caption{Example potential giving rise to $T_{RS}=3.6$, and the corresponding evolution in the $\phi$--$\chi$ plane.}
\label{fig:largetrs}
\end{figure}

Moving on to the results of the calculations done in the previous section, we look at the role played by isocurvature perturbations in modifying the adiabatic perturbations from their horizon-crossing value. Figure~\ref{fig:powerstransfers} confirms that the curvature power spectrum always increases, and for a non-negligible proportion of the time the fractional change is very large. Looking at the histograms for the transfer functions we see that the isocurvature perturbations are an important source, since at horizon exit they are of the same order as the curvature power spectrum and it is not particularly uncommon for $T_{RS}$ to be significant or even greater than one.

One consequence of the super-horizon fuelling of adiabatic perturbations is  that we have explicit violation of the consistency relation, as demonstrated in the plot of $r$ against $\epsilon$ in Fig.~\ref{fig:repsilon} (the tensor spectral index in these models is given by $n_{\rm T} = -2\epsilon$ as usual). Substituting eqs.~\eqref{eq:tensorpower} and \eqref{eq:PR} into eq.~\eqref{eq:r}, we have
\begin{equation}\label{eq:rofetrs}
r=\frac{16\epsilon}{1+4CT_{RS}\phi_{\perp}''/v+T_{RS}^2}.
\end{equation}
So we see we have a line of points corresponding to $r=16\epsilon$ as in the single scalar field case but with many points dropping below the line where $T_{RS}$ and the turn rate have suppressed the value of $r$. Hence the relation becomes an inequality when there are extra degrees of freedom, as first described in ref.~\cite{Sasaki}.

As an example of the sort of situation that can give rise to large $T_{RS}$ values we include Fig.~\ref{fig:largetrs}. As mentioned earlier, generally the dominant effect is the turn rate and while there are a number of types of trajectory that lead to large $T_{RS}$, in this case we see that it is the cumulative effect of a fairly continuously curved trajectory spiralling its way to a minimum.

\begin{figure}[t]
\centering
\includegraphics[width=14cm]{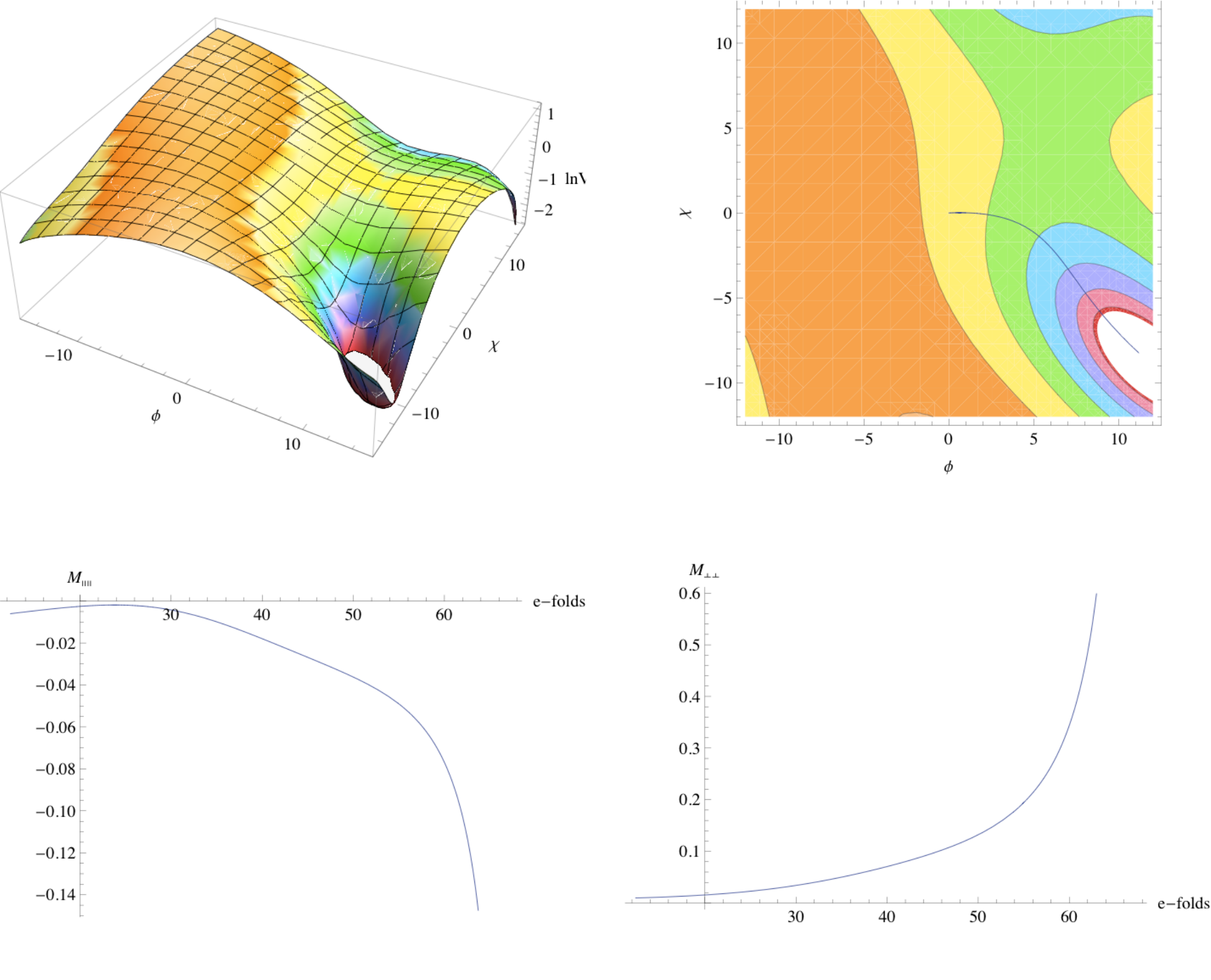}
\caption{Example plot of $\ln V$, the trajectory, and the relevant mass matrix components, showing why $T_{SS}$ tends to be so small.}
\label{fig:tssmcomps}
\end{figure}

As pointed out in ref.~\cite{Burgess} (see refs.~\cite{iso,WMAP7} for primary references), the CMB places strong observational constraints against the existence of isocurvature perturbations. As such, any multi-field model must either generate no isocurvature perturbations at horizon exit, or must find a way to make them disappear after horizon exit but before horizon entry. Our model falls nicely into the second category. Fig.~\ref{fig:powerstransfers} shows that the transfer function $T_{SS}$ consistently causes the isocurvature power spectrum to go to zero after horizon exit (except in one case which we will discuss in a moment). Referring back to eq.~\eqref{eq:TSS} and Table \ref{tab:isoevol}, we see that pure damping occurs when $M_{\parallel\parallel}<0$ and $M_{\perp\perp}>0$; we find that this scenario is typically the case, an example of which is shown in Fig.~\ref{fig:tssmcomps}. This result is not particularly surprising since one consequence of our choice of measure is that hilltop inflation is a very rare process, thus most inflation scenarios occur in the vicinity of a minimum. In such cases, at least to some extent the trajectory will be moving towards that minimum. We then have that perpendicular curvature will usually be positive and since the minimum must be very close to $V=0$, the log of the potential will have negative curvature in the direction of motion.  

\subsection{The curious case of Universe 1832942}

\begin{figure}[t]
\centering
\includegraphics[width=14cm]{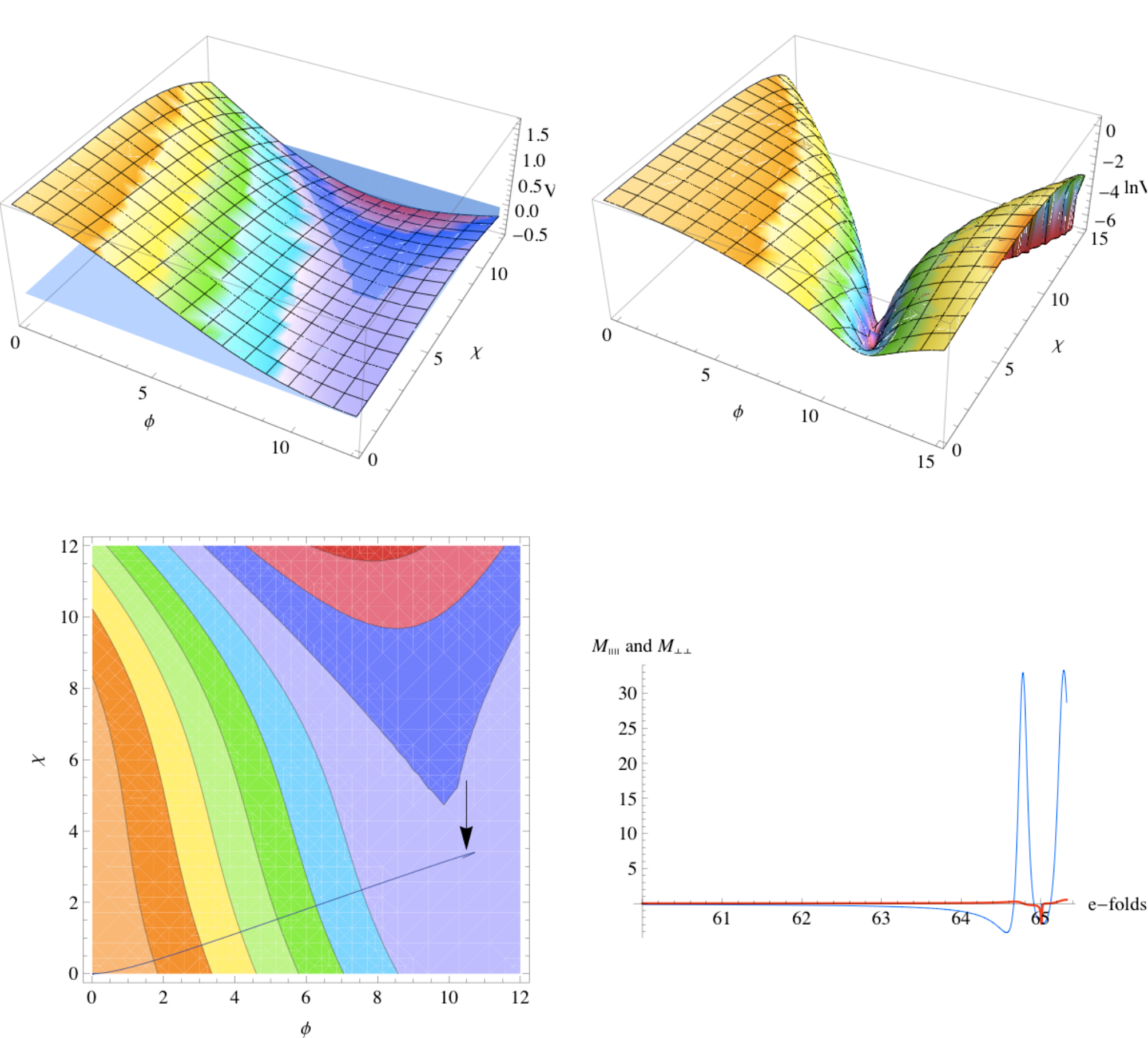}
\caption{Plots of the potential, log of the potential, trajectory and relevant components of the mass matrix ($M_{\parallel\parallel}$ is blue, $M_{\perp\perp}$ is red) for the one case we found, Universe 1832942, where the isocurvature perturbations were fuelled rather than going to zero. The arrow indicates where the trajectory turns back on itself.}
\label{fig:isofuel}
\end{figure}

Figure~\ref{fig:powerstransfers} shows one dramatic exception to the damping of the isocurvature perturbations just described. In one case the isocurvature perturbations were actually fuelled rather than damped. Fig.~\ref{fig:isofuel} shows plots of the potential, its logarithm, the trajectory, and a plot of the parallel and perpendicular components of the mass matrix near the end of inflation for this case which we call Universe 1832942. Remembering Table \ref{tab:isoevol}, we see that for the majority of the evolution the isocurvature perturbations are damped but near the end of the evolution the trajectory overshoots the minimum causing a sharp reversal in direction as it reaches the far side of the valley. As the trajectory goes through this sharp turn at low $V$ the arguments above reverse, but we see that the dominant effect in this case seems to be the change in curvature as the trajectory traverses the minimum. 

We have severe violation of SRST so our calculations of observables are no longer reliable, but we find the values $n_{\rm s}=0.94$ and $r=0.005$, with the low $r$ value resulting from a significant evolution of the power spectrum after horizon exit of $108\%$. We see that Universe 1832942 sits within the 95\% confidence contour of Fig.~\ref{fig:indexplots}, but we remind the reader that this confidence limit does not take surviving isocurvature modes into consideration.

\subsection{How will observables change with $D$?}

\begin{figure}
\centering
\includegraphics[width=15cm]{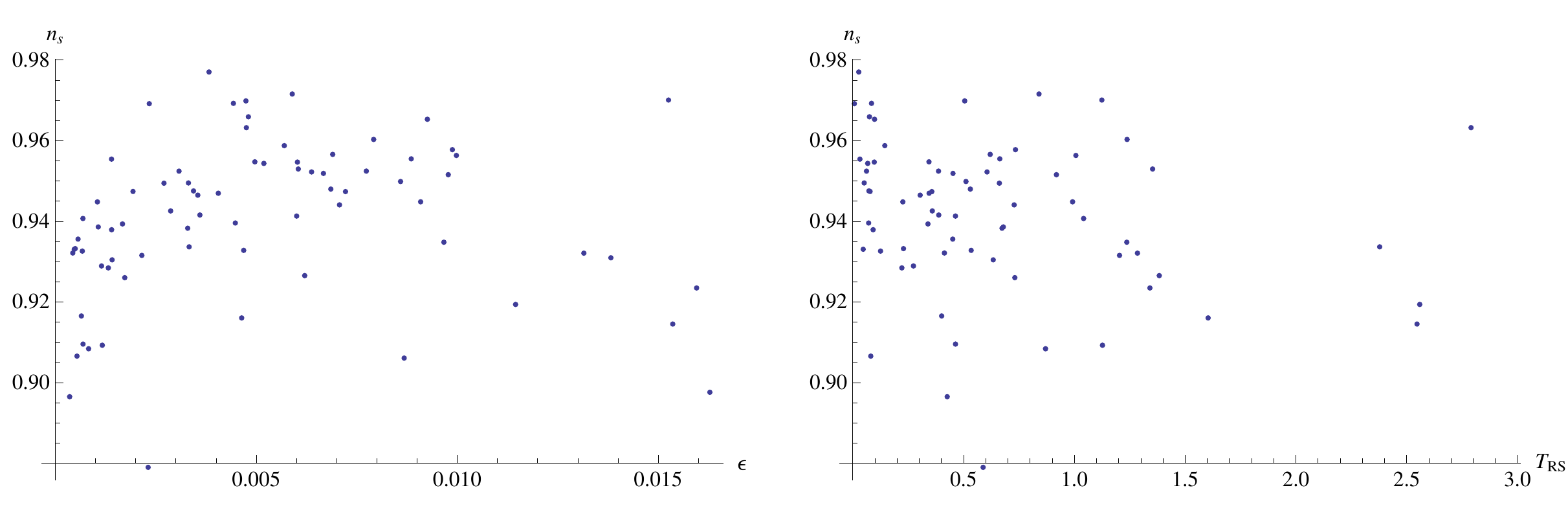}
\caption{Correlations of $n_{\rm{s}}$ with $\epsilon$ and $T_{RS}$.}
\label{fig:correl}
\end{figure}

As was mentioned at the beginning of this paper, we feel much of the phenomenology of a $D$ scalar field landscape is manifest with just two scalar fields. This is  true in terms of the types of phenomena that exist, but can we make any comments on how the values of observables might change with $D$? 

The short answer is no. From a statistical standpoint we know that the mean values of $\epsilon$ and $T_{RS}$ will both increase with $D$, so one might think that parameters such as $n_{\rm{s}}$ and $r$ could tend to some value as a result of one parameter coming to dominate. This may well be true but how so is not obvious. The lack of any strong correlation in Fig.~\ref{fig:correl} for $n_{\rm{s}}$ with respect to $\epsilon$ or $T_{RS}$ leads us to think that predicting how $n_{\rm{s}}$ will change with $D$ is non-trivial. That said, there is an important distinction between how $\epsilon$ and $T_{RS}$ will change with $D$. $T_{RS}$ is unbounded, in that as you increase $D$ you just get an increasing sum of $T_{RS_{i}}$ terms; this has both a cumulative effect and increases the statistical chance of encountering a large $T_{RS_{i}}$. The value of $\epsilon$ in contrast is capped, if for no other reason because inflation will end if $\epsilon$ is too large, but it may not ever even increase that much. 

There are a number of examples in the literature where a large number of scalar fields can actually lead to a decrease in $\epsilon$, one example being assisted inflation \cite{assisted}. So for cases where the absolute values of $\epsilon$ and $T_{RS}$ are important we can hazard a guess and say that for sufficiently large $D$ eventually the dominant effect will be $T_{RS}$. This is the case for the tensor-to-scalar ratio and remembering Fig.~\ref{fig:repsilon} and eq.~\eqref{eq:rofetrs} we would thus expect increasingly severe violation of the consistency relation, with a suppression of $r$ as we move to higher $D$. We will investigate this as part of a future study of the $D$-dimensional case.

\section{Discussion}

We have investigated the properties of inflationary trajectories in a toy-model landscape with two scalar fields, in which we fully track the effects of isocurvature perturbations. We have focussed on the well-motivated case where the variations in the potential correspond to masses of the order of $M_{\rm Pl}$.
Trajectories with sufficient inflation are rare, with one successful run per roughly $10^5$ randomly generated potentials, but the successful runs are typically in good accord with observational constraints on the perturbations, mostly lying within the region of the $n_{\rm s}$--$r$ plane delineated by the WMAP+BAO+$H_0$ 95\% confidence contour.

We find that isocurvature perturbations naturally go to zero after horizon exit. This is a direct result of the typical geometry of the landscape in the vicinity of an inflationary trajectory. For isocurvature perturbations to not go to zero one requires an unstable trajectory, which happened only in one of our successful realizations. Nevertheless, in many cases the isocurvature perturbations have a lasting consequence as we find significant fueling of the adiabatic perturbations from the entropy perturbations to be a fairly common occurrence. Because of this, we find that the tensor-to-scalar amplitude often lies below the level that would be predicted by the single-field consistency equation. One might speculate that this effect is likely to become more prominent as the number of scalar fields is increased, both from a statistical standpoint and as a purely cumulative effect, since such evolution is closely linked with turns in the trajectory. We plan to investigate this dependence in future work.

While our work significantly extends previous comparable analyses in the literature, such as ref.~\cite{TegmarkInf}, it nevertheless remains rudimentary and many further steps are needed if it is to become more realistic, beyond the issue of the number of scalar fields. Most pressing is a proper treatment of the way the measure problem affects our work, particularly in combination with anthropic selection. At the moment we have largely ignored this and obtained results which are based on a uniform distribution of initial conditions, rather than as drawn from random late-time observers. 

\acknowledgments
The authors were supported by the Science and Technology Facilities Council [grant numbers ST/1506029/1 and ST/F002858/1]. We thank Mafalda Dias for numerous discussions relating to this work, Courtney Peterson and Max Tegmark for clarifications relating to ref.~\cite{PTtwofield}, and Michael Salem and David Wands for very helpful suggestions.


\end{document}